\documentclass[letterpaper]{article}
\usepackage{aaai}
\usepackage[colorlinks=true, linkcolor=blue, citecolor=blue, urlcolor=blue]{hyperref}
\usepackage[numbers]{natbib}  

\usepackage{times}
\usepackage{float}
\usepackage{helvet}
\usepackage{courier}
\usepackage{dblfloatfix}
\usepackage[utf8]{inputenc} 
\usepackage[T1]{fontenc}    
\usepackage{url}            
\usepackage{booktabs}       
\usepackage{amsfonts}       
\usepackage{nicefrac}       
\usepackage{microtype}      
\usepackage{xcolor}         
\usepackage{microtype}      
\usepackage{lipsum}
\usepackage{graphicx, multirow}
\graphicspath{ {./images/} }
\usepackage{enumitem}
\usepackage{upgreek}
\usepackage{xcolor}         
\usepackage{enumitem}
\usepackage{color, colortbl}
\usepackage[dvipsnames]{xcolor}
\usepackage{amsmath}
\usepackage{amssymb}
\usepackage{wrapfig}
\usepackage{pifont}
\usepackage[font=small,labelfont=bf]{caption}

\newcommand{\rev}[1]{{\color{black} #1}} 

\title{{Hallucination Detection in Virtually-Stained\\Histology: A Latent Space Baseline}}
\author{
  \large{Ji-Hun Oh$^{1}$, Kianoush Falahkheirkhah$^{2}$, John Cheville$^{3}$, Rohit Bhargava$^{1,2,4\text{--}9}$} \\ \\
  \footnotesize{$^{1}$Department of Mechanical Science and Engineering, University of Illinois Urbana-Champaign, IL, US}\\
  \footnotesize{$^{2}$Beckman Institute for Advanced Science and Technology, University of Illinois Urbana-Champaign, IL, US}\\
  \footnotesize{$^{3}$Mayo Clinic, Rochester, MN, US}\\
  \footnotesize{$^{4}$Department of Bioengineering, University of Illinois Urbana-Champaign, IL, US}\\
  \footnotesize{$^{5}$Department of Electrical and Computer Engineering, University of Illinois Urbana-Champaign, IL, US}\\
  \footnotesize{$^{6}$Department of Chemical and Biomolecular Engineering, University of Illinois Urbana-Champaign, IL, US}\\
  \footnotesize{$^{7}$Department of Chemistry, University of Illinois Urbana-Champaign, IL, US}\\
  \footnotesize{$^{8}$Cancer Center at Illinois, University of Illinois Urbana-Champaign, IL, US}\\
  \footnotesize{$^{9}$CZ Biohub Chicago, LLC, Chicago, IL, US}
}

\begin{document}

\maketitle

\begin{abstract}
\textbf{Histopathologic analysis of stained tissue remains central to biomedical research and clinical care. Virtual staining (VS) offers a promising alternative, with potential to reduce costs and streamline workflows, yet hallucinations pose serious risks to clinical reliability. Here, we formalize the problem of hallucination detection in VS and propose a scalable post-hoc method: Neural Hallucination Precursor (NHP), which leverages the generator’s latent space to preemptively flag hallucinations. Extensive experiments across diverse VS tasks show NHP is both effective and robust. Critically, we also find that models with fewer hallucinations do not necessarily offer better detectability, exposing a gap in current VS evaluation and underscoring the need for hallucination detection benchmarks.}
\end{abstract}

\begin{figure*}[t]
  \centering
  \includegraphics[width=1\linewidth]{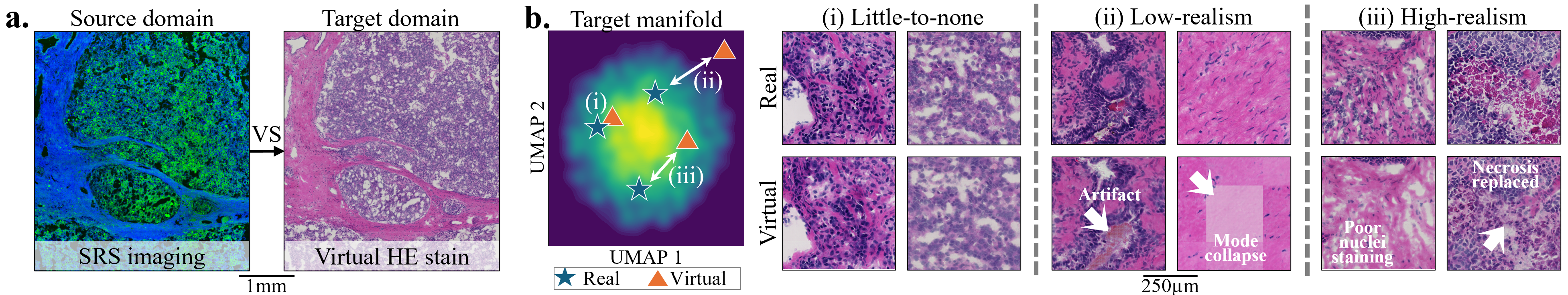}
    \caption{\rev{\textbf{a.} SRS$\rightarrow$H\&E VS task. \textbf{b.} Regions of interests (ROIs) with varying VS accuracy, with respect to the H\&E domain manifold: (i) Successful cases, where virtual H\&E closely matches the ground-truth H\&E; (ii) Low-realism hallucinations, clearly residing outside the target manifold; (iii) High-realism hallucinations, which deviate from the ground truth but lies within the manifold, making them difficult to detect.}}
   \label{fig1}
\end{figure*}

Histopathology relies on a century-old workflow: biopsies or surgical resections are removed from patients, preserved, sectioned, and stained, commonly with hematoxylin and eosin (H\&E), to highlight tissue structures for microscopic review by pathologists. This process, however, is labor-intensive, costly, and prone to artifacts and sample damage. Moreover, a single stain often lacks the full diagnostic context, prompting the use of multiple stains and exacerbating these issues. Recently, image-to-image translation (I2IT) has emerged as a compelling alternative, enabling computational generation of realistic stained images--an approach known as virtual staining (VS) \cite{liu2020global, de2021deep, liu2021unpaired, liu2022bci, zeng2022semi, lin2022unpaired, zhang2022mvfstain, boyd2022region, li2023adaptive, li2024virtual, chen2024pathological, wang2024mix, wei2024derestainer, liu2023vsgd, ma2023efficient, bayramoglu2017towards, schnell2020all, wolflein2023hoechstgan, he2024pst, guan2025supervised}. This covers tasks ranging from translating label-free modalities, such as autofluorescence (AF) or stimulated Raman scattering (SRS) imaging, to H\&E {(Fig. \ref{fig1}-a)}, as well as stain-to-stain conversion from routine H\&E to specialized immunohistochemistry (IHC) stains. This paradigm offers faster tissue assessments, at a lower cost and with simpler workflows, thereby streamlining patient care.

A critical challenge remains: {\textit{hallucinations}}. \rev{Fig. \ref{fig1}-b displays generated SRS$\rightarrow$H\&E images, where some images closely resemble the ground-truth H\&E stain (i), but others do not. Such false patterns display varying levels of realism, ranging from I2IT mode collapse (ii) to realistic patterns harder to manually distinguish, like histological features attenuated or substituted for others (iii).} False histology can mislead diagnosis or prognosis, risking adverse outcomes. This underscores the need for hallucination detection to facilitate safe, open-world VS deployment. Yet, this task has received limited attention, with only two recent studies \cite{huang2025robust, ounissi2025scalable} addressing it. In \S\ref{sec:2}, we outline the shortcomings in detection validation (limited scope or evaluation protocols) and methods (scalability or robustness challenges). The former is crucial, as even small errors in practice can distort detection evaluations; the latter is essential because VS involves massive terabyte whole slide image (WSI) datasets and can vary widely in tasks, requiring methods that are high-throughput, robust, and versatile.

Here, we address the first issue by formally defining the hallucination detection problem and its evaluation procedure (\S\ref{sec:3}). Specifically, we postulate the underlying causes of hallucinations, clarifying that detection is neither an out-of-distribution (OOD) nor outlier detection task, but one that must align with the VS prediction objective. To address the second issue, we hypothesize that hallucination precursors exist in the VS generator's latent space, enabling a simple baseline for hallucination detection. Termed \textit{Neural Hallucination Precursor} (NHP, \S\ref{sec:4}), we perform a post-hoc construction of a hallucination marker by combining feature signals and optimizing it for the VS task at hand. This meets key requirements for scalability, robustness, and versatility. In \S\ref{sec:5}, we demonstrate that NHP performs consistently across different organs, modality pairs, and I2IT backbones. Along the way, we further introduce new research themes and insights, particularly the disconnect between hallucination robustness and detectability--i.e., models with fewer hallucinations do not necessarily enable better detection, motivating the need for hallucination detection benchmarks. Overall, this work serves as a primer on hallucination detection in VS, establishing the problem, proposing a baseline, and outlining remaining challenges to guide future work.

\section{Related Work} 
\label{sec:2}
\subsection{VS Learning Paradigms} I2IT in VS can be either supervised by using source/target pairs from multiplexing techniques or label-free modalities \cite{bayramoglu2017towards, schnell2020all, wolflein2023hoechstgan, de2021deep}, or unsupervised by using unpaired sets \cite{zhang2022mvfstain, zeng2022semi, boyd2022region, liu2021unpaired, lin2022unpaired}. Supervised methods perform better but may not be a viable option due to technical overhead. An emerging trend is to use axially adjacent tissue slices as source/target pairs, meticulously registered \cite{li2024virtual, wang2024mix, wei2024derestainer, liu2022bci, li2023adaptive, chen2024pathological, he2024pst, guan2025supervised}. While not strictly paired, their proximity permits perceptual similarity, sometimes termed an ``inconsistent pair,'' which is generally accepted for full-reference (FR) metric evaluation, as well as for weak supervision during model training. Overall, generative adversarial networks (GANs) \cite{goodfellow2014generative} are by far the most widespread and mature I2IT backbone in VS, for both supervised and unsupervised approaches; we thus focus on GAN-based VS, in line with this trend.

\subsection{VS Hallucination Detection} Ref. \cite{huang2025robust} performed cyclic translations between AF (source) and H\&E (target) domains to define a hallucination index for lung and kidney biopsies. However, this incurs latency from iterative translations and is limited to GANs with the inverse target-to-source mapping model, requiring its additional training in cases where it is unavailable. Another study \cite{ounissi2025scalable} used the discriminator to monitor H\&E-to-IHC VS confidence in pediatric Crohn’s disease. These studies adopt detection principles from the GAN-based OOD/anomaly detection literature \cite{zenati2018adversarially, sabokrou2018adversarially, schlegl2019f}, but, as later discussed, hallucinations are not necessarily OOD (and vice versa), potentially restricting their detection scope. Furthermore, the validation of these studies is primarily visual or limited. For instance, \cite{huang2025robust} distinguished between ``good'' and ``poor'' VS models based on training status; while hallucinations are indeed more common in the latter, even well-trained models are not immune to them. Similarly, flagging corrupted source images, as done in \cite{ounissi2025scalable}, may not reliably indicate hallucinations when the VS model is robust. Recognizing these limitations, we aim to propose a universal, robust, and rigorously validated hallucination detection method.

\subsection{Detection vs. Mitigation} To date, mainstream VS research has primarily focused on mitigating hallucinations, e.g., by encouraging pathological semantics preservation \cite{liu2021unpaired, zeng2022semi, lin2022unpaired,boyd2022region, li2023adaptive, ma2023efficient, li2024virtual, chen2024pathological, wang2024mix, wei2024derestainer, he2024pst, guan2025supervised}. In contrast, the goal of this study is to detect hallucinations, {irrespective of their occurrence frequency}. These goals can be mutually inclusive, but our primary focus here is on detection. Nevertheless, we explore the relation between these two safety tasks later.

\begin{figure*}[t]
  \centering
  \includegraphics[width=1\linewidth]{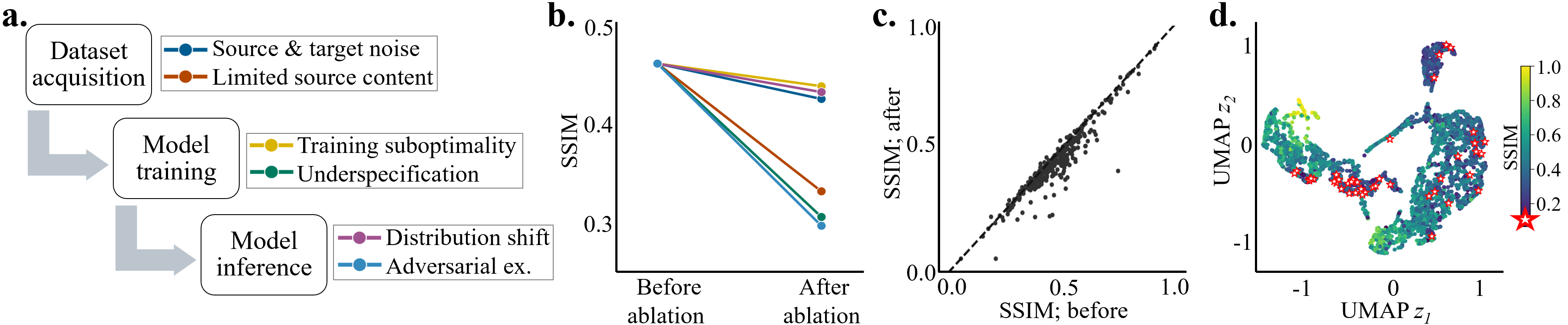}
    \caption{\textbf{a-b.} Hallucination causes across the VS pipeline, validated via targeted ablations in our Pix2PixHD SRS$\rightarrow$H\&E VS setup (\S\ref{sec:5}) and quantifying test set's $\Delta\mathcal{Q}$, specifically SSIM. (1) Data acquisition: Measurement noise/variability (additive SRS noise applied) and insufficient content (halving SRS spectral bands). (2) Model training: Suboptimal convergence (training prematurely halted) and underspecification (Pix2PixHD replaced with CycleGAN). (3) Model inference: Distribution shift (test samples OOD-shifted) and attacks (test samples adversarially manipulated). \textbf{c.} Sample-wise SSIM comparison before vs. after OOD-shifted. \textbf{c.} UMAP \cite{mcinnes2018umap} of VS model latent embeddings for ID data, color-coded by corresponding $\mathcal{Q}$.}
   \label{fig2}
\end{figure*}

\section{Problem Formulation}
\label{sec:3}
\textit{Setup:} Let $G^*: S \rightarrow T$ represent the underlying VS mapping function between source and target domains $S$ and $T$, consisting of WSI patches. Let $\mathcal{D}$ denote a training set with marginal distributions $P_S$ and $P_T$ for $S$ and $T$. Through empirical risk minimization (ERM), we estimate ${G} = {\arg\min}_{{G} \in H} \mathcal{L}({G}, \mathcal{D})$, where $\mathcal{L}$ is the Nash equilibrium loss of the GAN, and $H$ is the hypothesis space. Auxiliary notations like the discriminator are dropped. Afterwards, for test patch $\textbf{s}\in S$, inference is performed via $G(\textbf{s})$.

\subsection{Hallucination Hypothesis} Broadly defined, a ``hallucination'' occurs when a prediction $G(\mathbf{s})$ diverges from its corresponding ground-truth target image $\mathbf{t}$. This divergence can be operationalized via FR similarity metrics, denoted as $\mathcal{Q}$, where a lower $\mathcal{Q}(G(\mathbf{s}), \mathbf{t})$ signifies a greater extent of hallucination. The use of FR metrics is essential to detect realistic hallucinations--instances where $G(\mathbf{s})$ resides within the target distribution $P_T$ but fails to match $\mathbf{t}$--which no-reference metrics cannot capture. Importantly, this limits hallucination detection evaluation to fully, or at least inconsistently, paired datasets, both denoting reference target by $\textbf{t}$ for notation simplicity.

\rev{While the specific formulation of $\mathcal{Q}$ is context-dependent and inherently subjective, standard signal measures such as peak signal-to-noise ratio (PSNR) and structural similarity index (SSIM) remain prevalent for general quality control. Conversely, metrics aligned with clinical utility and pathologist perception offer superior assistive value in specialized domains. Here, we do not advocate for a specific hallucination metric; rather, our objective is to blindly estimate when a prescribed metric will indicate a hallucination. While we consider a wide range of metrics, including those clinically motivated, we default to PSNR, multi-scale SSIM, and learned perceptual image patch similarity (LPIPS; utilizing its negative value for directional consistency) \cite{zhang2018unreasonable} unless otherwise specified. These choices are consistent with established VS literature \cite{bayramoglu2017towards, zhang2022mvfstain, liu2022bci, liu2021unpaired, ma2023efficient, zeng2022semi, chen2024pathological, wei2024derestainer, li2023adaptive, liu2023vsgd, he2024pst, li2024virtual, guan2025supervised} while remaining computationally efficient.}

\subsection{Hallucination Causes} We group and analyze by their origin within the VS train-to-deploy pipeline:
\begin{enumerate}
    \item {Dataset acquisition:} $G^*$ is inherently ill-posed for ambiguous VS tasks due to measurement noise or insufficient source context \cite{stuart2010inverse}. This ill-posedness leads to one-to-many mappings, requiring a probabilistic formulation to faithfully approximate $G^*$. However, VS models are often deterministic by design; thus, a statistically valid ERM solution may yield a prediction $G(\mathbf{s})$ that deviates from the true target observation.
    \item {Model training:} Even when $G^*$ is well-posed, ${G} \neq G^*$ due to practical training challenges \cite{mescheder2018numerics, thanh2020catastrophic}, compounded by factors such as class imbalance \cite{cohen2018distribution} and domain asymmetry \cite{patashnik2021balagan}. More fundamentally, underspecification \cite{d2022underspecification} implies the existence of a Rashomon set \cite{breiman2001statistical}, $H^\dagger \subset H$, comprising infinitely many models that satisfy ERM equally well--i.e., $\mathcal{L}_{\text{val}}({G}, \mathcal{D}) < \tau$, $\forall {G} \in H^\dagger$, where $\mathcal{L}_{\text{val}}$ is a validation criterion and $\tau$ a performance threshold. The realized solution from $H^\dagger$ may not match $G^*$ as there is no explicit bias towards it. Underspecification is more acute in small datasets or under-constrained I2IT setups, such as unsupervised ones \cite{lu2019guiding, shen2020one}.
    \item {Model inference:} Training and test distributions are non-i.i.d., reflecting the high degree of heterogeneity in digital pathology data. This results in OOD encounters where $\textbf{s} \notin P_S$ (cf., {in-distribution}, ID, where $\textbf{s} \in P_S$). Common OOD sources include institutional or demographic shifts \cite{howard2021impact}, or unseen artifacts and pathologies \cite{oh2024we}. VS models are prone to hallucinate under OOD conditions, a corollary of underspecification: ERM constrains only ID behavior, allowing many $H^\dagger$ models to generalize poorly beyond it \cite{d2022underspecification, teney2022predicting}. Moreover, I2IT models lack adversarial robustness, with research showing that imperceptible perturbations can disrupt malicious I2IT systems like deepfake \cite{yeh2020disrupting} and watermark removers \cite{liu2022watermark}. By extrapolation, VS models may also be vulnerable to targeted hallucination attacks.
\end{enumerate}
These factors are demonstrated in Fig. \ref{fig2}-a-b where targeted hallucinogenic ablations to the VS pipeline deteriorates $\mathcal{Q}$. In literature, uncertainty is often categorized as aleatoric or epistemic \cite{hullermeier2021aleatoric}. Under this dichotomy, 1 is aleatoric, whereas 2-3 are epistemic.

\subsection{Detection Objective \& Evaluation} The goal is to devise a monitor, $f: S \rightarrow \mathbb{R}$, that blindly predicts input-specific VS confidence, i.e., $\mathcal{Q}$. To evaluate performance, we conduct an abstention test \cite{malininensemble, van2020uncertainty, postels2022practicality}: Given a test set $\mathcal{D}_{\text{test}}$, we set threshold $\lambda_{p}$ to reject top-$p\%$ of predicted hallucinatory samples, satisfying $\text{Pr}_{(\textbf{s},\textbf{t}) \in \mathcal{D}_{\text{test}}}[f(\textbf{s}) \geqslant \lambda_p] = 1-p$, and compute $\mathcal{Q}$ over the retained predictions. To sidestep the sensitivity of $p$, we sweep $p\in[0,1]$ and measure the area-under-curve (AUC):
\begin{equation}
AUC_{f} = \int^{1}_{0} \mathbb{E}_{(\textbf{s},\textbf{t}) \in \mathcal{D}_{\text{test}} \, | \, f(\textbf{s}) \geqslant \lambda_p} \bigr[ \mathcal{Q}({G}(\textbf{s}), \textbf{t})\bigr] dp.
\label{eq1}
\end{equation}
As this is biased toward the base performance of the VS model, we adjust via normalization w.r.t. two baselines: a random monitor, with $AUC_{\text{random}}=\mathbb{E}_{(\mathbf{s}, \mathbf{t}) \in \mathcal{D}_{\text{test}}}[\mathcal{Q}(G(\mathbf{s}), \mathbf{t})]$, and an oracle, which uses $\mathcal{Q}(G(\mathbf{s}), \mathbf{t})$ itself as the confidence score, yielding an upper bound $AUC_{\text{oracle}}$. The final metric, referred to as \textit{{Hallucination Rejection Preference (HRP)}}, is:
\begin{equation}
HRP = \frac{(AUC_f - AUC_{\text{random}})}{(AUC_{\text{oracle}} - AUC_{\text{random}})}.
\label{eq2}
\end{equation}
Higher HRP (approaching 1) indicates better monitor performance, computed per $\mathcal{Q}$ metric. 

\subsection{Clarifications: OOD vs. Hallucination Detection} Hallucination detection is not OOD detection, as OOD status does not inherently imply generalization failure. As shown in Fig. \ref{fig2}-b-c, our VS models exhibit robustness with only marginal $\mathcal{Q}$ degradation and few sample failures; to maintain open-world utility and avoid excessive false alarms, only fatal instances should be rejected rather than all OOD cases. Furthermore, hallucinations are not exclusive to OOD data; ID samples are susceptible due to domain-agnostic factors such as data ambiguity or underspecification. Thus, OOD status is neither necessary nor sufficient for hallucination. While recent classification literature \cite{guerin2023out, jaegercall, averly2023unified} also advocates for ID/OOD-agnostic error detection, VS differs fundamentally regarding semantic (novelty) OOD. Unlike closed-set classification, which must reject novel labels by design, VS is unconstrained and can accurately synthesize novel data types \cite{tonks2024can}. Consequently, semantic OODs in VS should be treated as functional samples rather than automatic failures.

Another point of confusion is mistaking this as an outlier problem, assuming that only outlying IDs hallucinate. However, as discussed, many hallucination causes are distribution-agnostic, refuting this notion. We show this in Fig. \ref{fig2}-d, {where not all outliers hallucinate, and vice versa}. All this leads to the conclusion that distribution-based detection tasks (OOD, novelty, anomaly, outlier) are unideal proxies for hallucination detection. The same holds for other hallucination factors as well: hallucination detection should not solely focus on, e.g., ambiguity or underspecification. Instead, it must align with VS predictions--{detecting when (and only when) $\mathcal{Q}$ is poor}.

\section{Method}
\label{sec:4}
\subsection{NHP Formulation}\label{sec:4.1}
\begin{figure}[t]
  \centering
  \includegraphics[width=\linewidth]{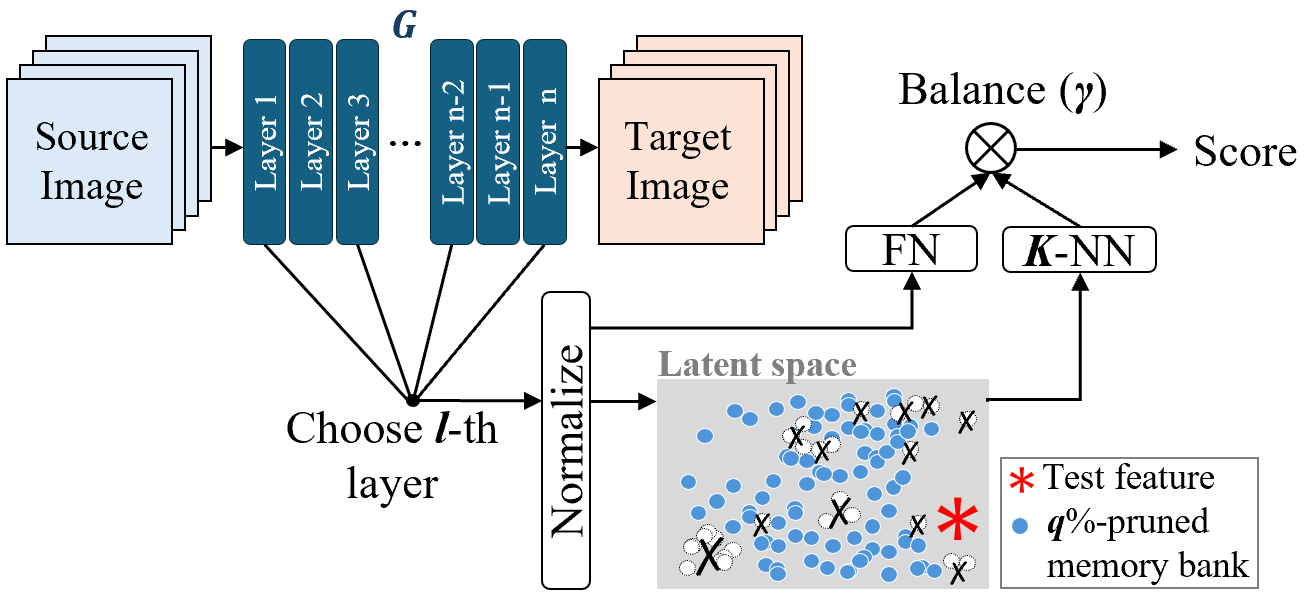}
    \caption{Schematic of NHP.}
   \label{fig3}
\end{figure}

We introduce NHP, a baseline method for detecting VS hallucinations by identifying statistical deviations from a feature memory bank within the generator’s latent space (Fig. \ref{fig3}). This bank is constructed using a fully or inconsistently paired calibration set, $\mathcal{D}_c$, typically available through clinical validation protocols or, if applicable, sampled from the training set. However, directly using $\mathcal{D}_c$ is problematic, as latent hallucinations within the set could lead to the erroneous association of unsafe features as ``safe.'' To address this, we prune the top-$q$\% of hallucinations based on prescribed $\mathcal{Q}$ metrics:
\begin{equation}
\mathcal{D}^q_c = \bigl\{ (\textbf{s}_c, \textbf{t}_c) \in \mathcal{D}_c \mid \mathcal{Q}({G}(\textbf{s}_c), \textbf{t}_c) \geqslant \lambda_{\mathcal{Q}}^q, \forall \mathcal{Q} \bigl\},
\label{eq4}
\end{equation}
where the threshold $\lambda_{\mathcal{Q}}^q$ satisfies $\text{Pr}[\mathcal{Q}({G}(\textbf{s}_c), \textbf{t}_c) \geqslant \lambda_{\mathcal{Q}}^q] = 1-q$. Subsequently, we extract and spatially average the $l$-th layer feature blocks:
\begin{equation}
\textbf{z}_c^l = AvgPool(G^l(\textbf{s}_c)),\quad\forall\textbf{s}_c \in \mathcal{D}_c^q
\label{eq3}
\end{equation}
resulting in a feature memory bank, $Z^q_c \subseteq \mathbb{R}^{N \times D}$, where $N$ denotes the number of pruned calibration samples and $D$ represents the channel-wise dimensionality.

For a test image $\textbf{s}$, we extract the corresponding feature $\textbf{z}^l$ and measure its deviation from the bank via a generalized scoring function:
\begin{equation}
    f_{\text{NHP}}(\textbf{s}) = -r_{(k)} \cdot \|\textbf{z}^l\|_2^{\gamma},
\label{eq5}
\end{equation}
where $r_{(k)}$ represents the normalized $k$-nearest neighbor (KNN) distance and $\gamma$ is a balancing coefficient for the $\ell_2$ feature norm (FN). Specifically, $r_{(k)}$ is defined as the $k$-th smallest $\ell_2$-distance between the normalized query feature $\textbf{z}^l/\|\textbf{z}^l\|_2$ and the bank entries $\textbf{z}^l_c/\|\textbf{z}^l_c\|_2, \forall \textbf{z}^l_c \in Z^q_c$. The negative sign ensures that larger distances denote lower confidence, i.e., higher hallucination threat.

NHP involves four hyperparameters: $l$, $q$, $k$, and $\gamma$. To optimize these, we hold out a portion of $\mathcal{D}_c$ for validation and perform a grid search to maximize HRP over the pre-scribed target $\mathcal{Q}$ metrics. Crucially, hallucinations are not pruned here to ensure the parameters are tuned for optimal detection performance. When $\mathcal{D}_c$ is subsampled from the training set, we define this process as ``self-tuning.'' We hypothesize that since every VS model is unique, no single parameter set is universally optimal; this fine-tuning adapts the monitor to specific VS contexts while maintaining a general-purpose framework applicable to any VS modality or I2IT backbone. \rev{While the specific case of self-tuning may seem counterintuitive, as the training set reflects some hallucinogenic factors like data ambiguity but not OOD, neural network representations often exhibit robustness to distribution shifts that do not degrade performance. We therefore hypothesize that non-hallucinatory OOD data remain proximal to the safe bank, whereas hallucinatory cases deviate, enabling separation.}

\subsection{Comparison with Existing Methods}\label{sec:4.2}
\subsubsection{Naive KNN} NHP builds upon KNN-based supervised OOD detection for image classification \cite{sun2022out}, introducing several essential modifications to account for the distinct contextual settings of I2IT (vs. classification) and hallucination detection (vs. OOD). (\textit{naive} $\Rightarrow$ \textit{modified}):
\begin{itemize}
    \item \textit{Penultimate layer $\Rightarrow$ $l$-tuning:} Deeper layers capture high-level semantics (cf. shallow layers encode low-level attributes like edges), motivating the use of bottleneck or penultimate layers in existing latent space methods. However, hallucinations, as broadly defined, may not always be semantic and vice versa, challenging this norm. Since the optimal layer is unknown, we tune $l$.
    \item \textit{No Pruning $\Rightarrow$ $q$-tuning:} Previous works typically use the training subset as the safe set ($q=0$). This is suitable for distinguishing ID from OOD, as the training set is unequivocally ``safe'' in this context. However, our focus is on hallucinations, requiring the explicit removal of unsafe ID hallucinations to reflect deviations relative to a hallucination-free ID, not just any ID instance.
    \item \textit{No FN $\Rightarrow$ $\gamma$-tuning:} ID data typically exhibit higher FN values \cite{dhamija2018reducing, tack2020csi, sun2022out}, leading \cite{sun2022out} to initially decouple FN to prevent ID data from being far-distant, which would otherwise confound OOD detection. Apart from this, FN is not utilized ($\gamma=0$). However, this assumption does not hold in the context of I2IT. A recent study \cite{park2023understanding} links FN behavior to the maximum logit under regularity conditions, but this does not apply to non-classification tasks like I2IT--nor is it desirable, as ID data is not exclusively exempt from detection. While the precise interplay of FN remains unclear, it is an important source of information nonetheless, evident by its use in various context \cite{park2023understanding, xu2019larger, wei2023edge, meng2021magface}. Therefore, we retain FN with a granular balance through $\gamma$-tuning.
    \item \textit{Advanced/no tuning $\Rightarrow$ self-tuning}: In OOD literature, hyperparameters are often optimized through advanced schemes, such as using pseudo-OODs via jigsaw puzzles \cite{yu2023block} or adversarial examples \cite{lee2018simple}, which introduce complexity. Or, they are not tuned at all and (suboptimally) adopt default settings from other existing works. In our specific case of self-tuning, the presence of hallucinations in training avoids the necessity of such complexities.
\end{itemize}

\rev{\subsubsection{Latent Space Methods} NHP belongs to the broader class of latent space methods, also termed deterministic uncertainty quantification (DUQ) \cite{van2020uncertainty, mukhoti2023deep, postels2022practicality, sun2022out, lee2018simple, wang2022vim, henzinger2020outside, yang2023full}, where KNN serves as a measure of feature deviation. While alternative distance- or density-based measures, such as Gaussian models, principal subspaces, or graph-based methods, could theoretically substitute for KNN in Eq. \ref{eq5}, we found empirically that KNN yields the best performance.}

\rev{Rather than post-hoc fitting, one can explicitly model the generator’s latent space density via variational Bayes \cite{kingma2013auto} and perform likelihood-based scoring \cite{nalisnick2019deep,havtorn2021hierarchical}. While VAEs see limited use in VS due to lower image fidelity, the approach could be adapted to the NHP framework with similar contextual modifications. However, several challenges emerge: the latent variable must be fixed prior to training, yet the optimal layer is unknown, and pruning requires reshaping the latent distribution before likelihood estimation--both nontrivial tasks.}

\rev{\subsubsection{Unsupervised Methods} NHP operates within a supervised detection paradigm, where detection is calibrated to a specific model. In contrast, unsupervised methods, such as Variational Autoencoders (VAEs) that utilize likelihood-based or reconstruction error scoring \cite{nalisnick2019deep, havtorn2021hierarchical}, could be applied to either the source or target domain. However, because these methods focus on domain-centric OOD detection, they often misalign with the task-specific predictive confidence of the VS model; e.g., they fail to selectively target critical OOD instances or identify non-OOD hallucination sources.}

\subsection{Complexity}
\rev{NHP requires a KNN search during inference. In this work, we utilize Faiss’s \texttt{IndexFlatL2} \cite{johnson2019billion}, which scales with $\mathcal{O}(ND)$. While increasing the bank size ($N$) improves detector performance, it may impede real-time detection. However, we later show NHP remains effective with relatively small bank sizes ($N \sim 5$K). For a feature dimensionality of $D=64$ (typical for the penultimate layer of VS generators), this corresponds to sub-millisecond latency for a single patch on a commercial CPU. The memory bandwidth is only $\sim$1.28 MB (assuming float32 precision), fitting within L2/L3 caches; thus, with batch processing, this is significantly sped up to e.g., $\sim$1ms per 100 patches. NHP even works with aggressively fewer samples ($N \ll 1$K), further reducing these complexities. Overall, these times are orders of magnitude faster than the VS model’s GPU forward pass, making the computational overhead near negligible. Furthermore, as a post-hoc method, NHP requires no model modifications, additional training, or forward passes, unlike common uncertainty quantification methods like Monte Carlo Dropout \cite{gal2016dropout} and Deep Ensemble \cite{lakshminarayanan2017simple}. These properties ensure NHP is both scalable and straightforward for practitioners to implement.}

\section{Experiments}
\label{sec:5}
\subsection{Experimental Settings}
\label{sec:5.1}
\subsubsection{VS Datasets and Models} We evaluate NHP across seven VS tasks: SRS to H\&E in prostate cancer \cite{falahkheirkhah2023accelerating}; Hoechst 33342 (HO342) to immunofluorescence (IF) T-cell markers (CD3, CD8) in renal cancer \cite{wolflein2023hoechstgan}; and H\&E to four IHC stains (ER, HER2, PR, Ki67) in breast cancer \cite{li2023adaptive}. All tasks utilize tessellated 256$\times$256 pixel$^2$ patches, comprising approximately 4K, 404K, and 69K training samples for the SRS, HO342, and MIST datasets, respectively. For paired SRS and HO342 tasks, we train VS models using Pix2PixHD \cite{wang2018high} and a hybrid VSGD \cite{liu2023vsgd} + PatchNCE loss \cite{andonian2021contrastive} (VSGD+pNCE). For MIST, we assess CycleGAN \cite{zhu2017unpaired} and CUT \cite{park2020contrastive}. Although MIST is trained unsupervised, the dataset is inconsistently paired, allowing for FR metric computation and hallucination evaluation. This yields $7 \times 2$ configurations, each evaluated across 10 random seeds (140 VS models total). Implementations follow official repositories: batch sizes (8, 8, 1), learning rates (2e-3, 2e-4, 2e-4), and epochs (30, 1, 10) for SRS, HO342, and MIST, respectively. Evaluation uses external test subsets of approximately 500, 3K, and 2K patches. Overall, our setup spans diverse cancer types, modalities, training sizes (4K--475K), and I2IT backbones.

\begin{table*}[t]
\small
\setlength{\tabcolsep}{1.5pt}
\renewcommand{\arraystretch}{0.9}
\centering
\resizebox{1\textwidth}{!}{%
\begin{tabular}{l|cc|cc|cc|cc|cc|cc|cc|c}
\toprule
& \multicolumn{2}{c}{SRS$\rightarrow$H\&E}
& \multicolumn{2}{c}{HO342$\rightarrow$CD3}
& \multicolumn{2}{c}{HO342$\rightarrow$CD8}
& \multicolumn{2}{c}{HE$\rightarrow$ER}
& \multicolumn{2}{c}{HE$\rightarrow$HER2}
& \multicolumn{2}{c}{HE$\rightarrow$PR}
& \multicolumn{2}{c|}{HE$\rightarrow$Ki67} 
& \\
Method
& {\scriptsize Pix2PixHD} & {\scriptsize VSGD+pNCE}
& {\scriptsize Pix2PixHD} & {\scriptsize VSGD+pNCE}
& {\scriptsize Pix2PixHD} & {\scriptsize VSGD+pNCE}
& {\scriptsize CycleGAN} & {\scriptsize CUT}
& {\scriptsize CycleGAN} & {\scriptsize CUT}
& {\scriptsize CycleGAN} & {\scriptsize CUT}
& {\scriptsize CycleGAN} & {\scriptsize CUT} & \textbf{Avg.} \\
\midrule
ALOCC & -33.2\textsubscript{±10.3} & -0.4\textsubscript{±20.8} & -30.8\textsubscript{±4.8} & 15.6\textsubscript{±17.0} & -7.5\textsubscript{±9.1} & 15.7\textsubscript{±20.0} & 3.0\textsubscript{±44.2} & 0.6\textsubscript{±26.1} & 11.8\textsubscript{±24.4} & 3.2\textsubscript{±26.2} & 9.4\textsubscript{±46.3} & -25.9\textsubscript{±12.5} & 24.8\textsubscript{±20.4} & 4.0\textsubscript{±29.4} & -0.7 \\
ALAD & - & - & - & - & - & - & -6.8\textsubscript{±9.0} & - & -7.0\textsubscript{±15.0} & - & -3.9\textsubscript{±12.3} & - & 2.5\textsubscript{±22.8} & - & -3.8$^\dagger$ \\
f-AnoGAN & - & - & - & - & - & - & 3.9\textsubscript{±7.6} & - & -4.1\textsubscript{±7.1} & - & 4.8\textsubscript{±3.4} & - & 8.6\textsubscript{±5.9} & - & 3.3$^\dagger$ \\
\rev{DE, 4 mem.} & -5.8\textsubscript{±25.5} & 11.6\textsubscript{±27.6} & 19.6\textsubscript{±25.3} & 19.2\textsubscript{±21.8} & 20.2\textsubscript{±20.9} & 4.2\textsubscript{±6.3} & 32.1\textsubscript{±29.5} & 26.9\textsubscript{±19.4} & 11.8\textsubscript{±26.6} & 6.0\textsubscript{±16.3} & 18.3\textsubscript{±37.4} & 13.7\textsubscript{±31.0} & 4.6\textsubscript{±35.1} & 4.1\textsubscript{±18.8} & 13.3 \\
\rev{DE, 10 mem.} & 40.4\textsubscript{±3.2} & \textbf{47.5\textsubscript{±1.6}} & {51.2\textsubscript{±1.1}} & 36.0\textsubscript{±10.8} & 42.1\textsubscript{±3.1} & 8.1\textsubscript{±3.6} & 58.0\textsubscript{±1.9} & 43.7\textsubscript{±4.1} & 43.3\textsubscript{±4.3} & 31.9\textsubscript{±10.4} & 62.4\textsubscript{±2.0} & 47.2\textsubscript{±5.5} & 57.1\textsubscript{±2.0} & 24.5\textsubscript{±4.2} & 42.4 \\
NHP, $\gamma=0$ & 40.7\textsubscript{±8.5} & 36.7\textsubscript{±5.8} & 40.3\textsubscript{±3.4} & 49.3\textsubscript{±16.9} & 24.2\textsubscript{±5.3} & 30.7\textsubscript{±14.8} & 68.0\textsubscript{±3.1} & 56.4\textsubscript{±5.3} & 50.1\textsubscript{±5.1} & 42.8\textsubscript{±8.5} & \underline{68.6\textsubscript{±1.9}} & 54.0\textsubscript{±5.3} & 66.7\textsubscript{±3.5} & 40.9\textsubscript{±18.9} & 47.8 \\
NHP, $q=0$ & 21.4\textsubscript{±13.0} & 18.3\textsubscript{±10.8} & 50.5\textsubscript{±1.1} & 47.2\textsubscript{±15.2} & 43.6\textsubscript{±5.1} & 30.9\textsubscript{±13.9} & 47.7\textsubscript{±3.8} & 41.9\textsubscript{±14.6} & 42.2\textsubscript{±6.7} & 24.4\textsubscript{±7.8} & 57.4\textsubscript{±4.0} & 41.3\textsubscript{±10.3} & 57.4\textsubscript{±4.6} & 37.6\textsubscript{±22.0} & 40.1 \\
NHP, linear & 40.7\textsubscript{±8.5} & 36.7\textsubscript{±5.8} & 50.6\textsubscript{±4.8} & 49.4\textsubscript{±15.7} & 42.6\textsubscript{±8.0} & \underline{33.3\textsubscript{±16.0}} & \textbf{70.5\textsubscript{±2.6}} & \underline{61.6\textsubscript{±6.9}} & 50.6\textsubscript{±5.4} & 42.9\textsubscript{±9.3} & 68.6\textsubscript{±2.0} & \underline{57.2\textsubscript{±4.9}} & \textbf{67.1\textsubscript{±3.7}} & \textbf{50.4\textsubscript{±17.4}} & \underline{51.6} \\
NHP, OtB & 13.1\textsubscript{±9.3} & 34.8\textsubscript{±8.9} & 40.5\textsubscript{±5.6} & 39.9\textsubscript{±13.9} & 29.4\textsubscript{±9.7} & 27.9\textsubscript{±12.1} & 61.1\textsubscript{±2.2} & 49.5\textsubscript{±4.3} & 45.0\textsubscript{±4.3} & 35.1\textsubscript{±9.3} & 63.9\textsubscript{±2.6} & 51.5\textsubscript{±4.9} & 61.7\textsubscript{±3.1} & 38.7\textsubscript{±16.0} & 42.3 \\
NHP, Res. & 12.0\textsubscript{±17.0} & 2.8\textsubscript{±17.3} & \textbf{53.1\textsubscript{±1.5}} & 49.3\textsubscript{±16.2} & \textbf{47.7\textsubscript{±3.0}} & 32.5\textsubscript{±18.1} & 37.7\textsubscript{±7.5} & 30.1\textsubscript{±19.8} & 46.4\textsubscript{±7.6} & 22.4\textsubscript{±8.6} & 48.0\textsubscript{±6.4} & 31.7\textsubscript{±21.9} & 61.1\textsubscript{±5.4} & 47.5\textsubscript{±15.4} & 37.3 \\
NHP, GMM & \underline{49.0\textsubscript{±5.5}} & 28.7\textsubscript{±9.9} & 41.6\textsubscript{±4.0} & \textbf{49.5\textsubscript{±16.9}} & 33.6\textsubscript{±6.5} & 32.0\textsubscript{±14.0} & 66.7\textsubscript{±2.0} & 58.3\textsubscript{±5.1} & \underline{50.7\textsubscript{±5.3}} & \textbf{43.9\textsubscript{±8.2}} & 68.5\textsubscript{±2.4} & 54.8\textsubscript{±6.5} & 65.4\textsubscript{±3.1} & 42.2\textsubscript{±22.9} & 48.9 \\
NHP, \textit{ours} & \textbf{49.3\textsubscript{±5.5}} & \underline{37.2\textsubscript{±5.7}} & \underline{51.3\textsubscript{±4.6}} & \underline{49.5\textsubscript{±15.4}} & \underline{45.7\textsubscript{±6.1}} & \textbf{33.5\textsubscript{±15.9}} & \underline{69.3\textsubscript{±2.9}} & \textbf{62.2\textsubscript{±6.3}} & \textbf{51.0\textsubscript{±5.5}} & \underline{43.4\textsubscript{±8.9}} & \textbf{68.9\textsubscript{±1.9}} & \textbf{57.8\textsubscript{±4.5}} & \underline{67.0\textsubscript{±3.6}} & \underline{47.9\textsubscript{±20.5}} & \textbf{52.4} \\
\bottomrule
\end{tabular}}
\caption{Mean (±std.) HRP scores (\%, $\uparrow$) over $\mathcal{Q}$ metrics PSNR, SSIM, and LPIPS. Best method in \textbf{bold}, runner-up \underline{underlined}. $\dagger$ denotes average over available tasks.}
\label{tab1}
\end{table*}

\subsubsection{NHP Implementation} For the NHP grid search, we sweep: $l \in \{0, 0.25, 0.5, 0.75, 1\}$, representing the sequential layer index\footnote{\rev{A first-order proxy for feature hierarchy. While this is less precise in the presence of skip or residual connections, such precision is not critical as long as the search range encompasses sufficiently diverse feature representations.}} (0 = first, 1 = penultimate); $q \in \{0, 0.25, 0.5, 0.75\}$, ranging from no pruning ($q=0$) to aggressive pruning ($q=0.75$); $k \in \{1, 10, 25, 50, 100, 200\}$, spanning local to global structures; and $\gamma \in [-10, 10]$ in steps of 0.5, where higher values amplify ($\gamma \gg 0$), zero nullifies ($\gamma=0$), and lower values invert ($\gamma \ll 0$) the FN term. \rev{This broad search space is designed to capture a wide functional range; while prior task-specific knowledge could narrow these bounds, we evaluate NHP in an assumption-free setting.} For $\mathcal{D}_c$, we subsample from the training set (self-tune), as all datasets are at least inconsistently paired and this imposes a less restrictive setting. We utilize the full SRS training set and stratified subsets for HO342 (1\%) and MIST (10\%), yielding $\sim$3--7K samples per task. For each of the 140 VS models, we execute NHP 10 times using different 25\% self-tuning splits, totaling 1,400 NHP instances.

\subsubsection{Comparison Methods} We compare NHP against three categories of methods:
\begin{itemize}
    \item {GAN-based:} We evaluate GAN-based principles from recent VS hallucination literature \cite{ounissi2025scalable, huang2025robust}, specifically: ALOCC \cite{sabokrou2018adversarially, zaheer2020old}, using the target discriminator’s output as confidence; ALAD, using the source discriminator’s feature matching error between source and inverse reconstruction (source$\rightarrow$target$\rightarrow$source); and f-AnoGAN \cite{schlegl2019f}, which integrates ALAD with the source inverse reconstruction residual, though we evaluate only the latter for isolated assessment. Note that ALAD and f-AnoGAN apply only to CycleGANs.
    \item \rev{Deep ensemble (DE): Although not previously used in VS, we evaluate DE \cite{lakshminarayanan2017simple} as a gold-standard uncertainty method 
    \cite{mehrtens2023benchmarking}. Standard DE averages model outputs and uses the variance as uncertainty. In I2IT, to avoid blurring, we randomly select one ensemble prediction and define its confidence as the average of pairwise $\mathcal{Q}$ scores across all members. This is calculated for each $\mathcal{Q}$ metric, then standardized and averaged, capturing multi-metric divergence rather than simple pixel-wise variance. We evaluate two configurations: a practical 4-member setup (averaged across 10 trials) and a more intensive 10-member setup.}
    \item \rev{Latent space:} \color{black}We evaluate latent space methods without NHP modifications or using alternative formulations. Ablations include the naive KNN baseline \cite{sun2022out} and versions omitting the FN term ($\gamma=0$), omitting pruning ($q=0$), or adopting a linear balance in Eq. \ref{eq5} ($-r_{(k)} + \gamma \cdot \|\mathbf{z}^l\|_2$). We additionally evaluate three distance-based scores, standardized before merging with FN and similarly tuned via grid search: Outside-the-Box (OtB) \cite{henzinger2020outside}, which fits a hypercube based on the $p_{\in\{1, 0.99, 0.975, 0.95, 0.9, 0.8\}}$-th percentile of features in $Z^q_c$ and uses the in-box feature ratio of $\mathbf{z}^l$; Residual \cite{wang2022vim}, which projects $\mathbf{z}^l$ onto the orthogonal residual subspace of $Z^q_c$ (explained variance ratio $V_{\in\{0.01, 0.025, 0.05, 0.075, 0.1, 0.2\}}$) and uses the negative $\ell_1$-norm; and GMM \cite{yang2023full}, which fits a Gaussian mixture model via $Z^q_c$ with $C_{\in\{1,4,8,16,32,64\}}$ clusters and uses the log-likelihood of $\mathbf{z}^l$.
\end{itemize}

\begin{figure*}[t]
  \centering
  \includegraphics[width=\linewidth]{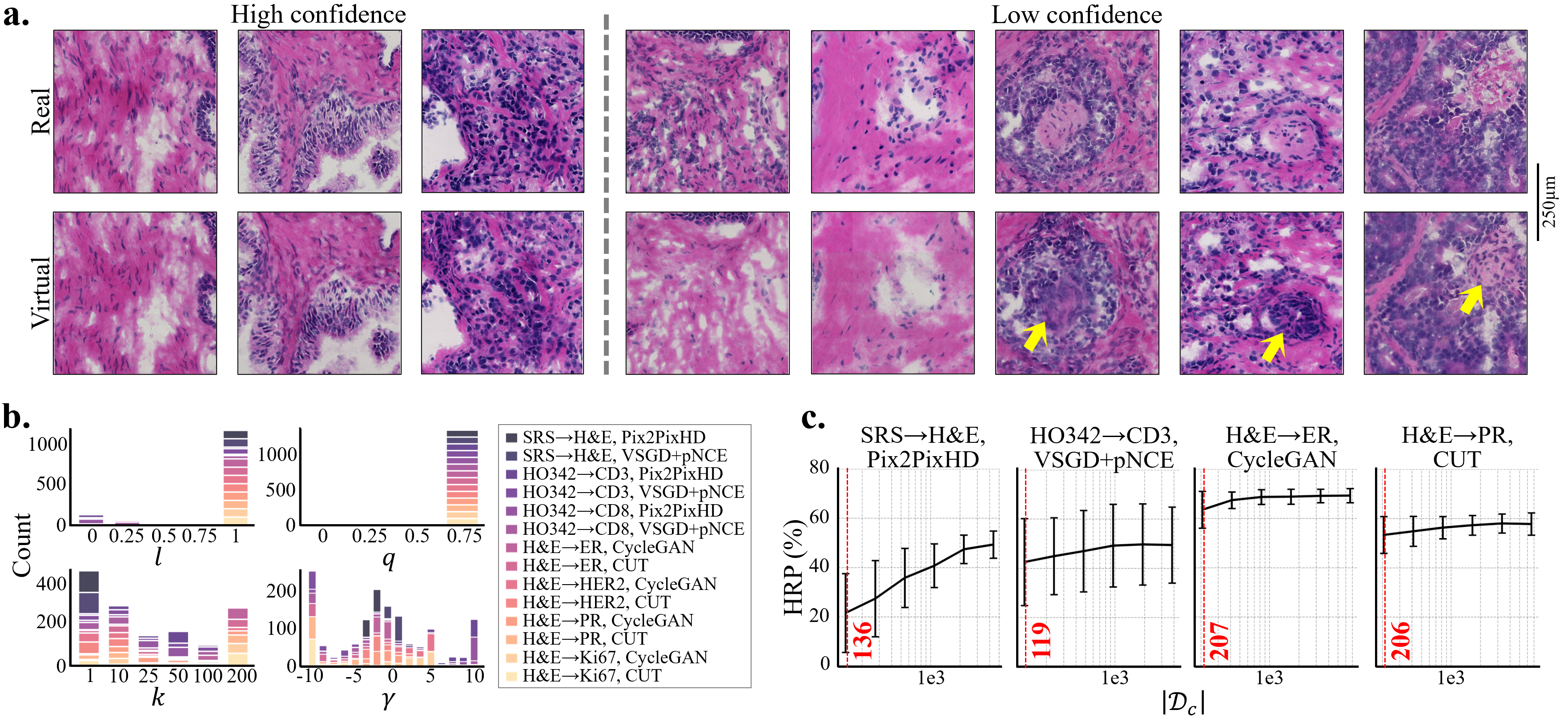}
   \caption{\rev{\textbf{a.} Example ROIs of high vs. low NHP confidence in the SRS$\rightarrow$H\&E task.} \textbf{b.} Converged NHP hyperparameters, aggregated from 1400 NHPs. \textbf{c.} Mean HRP using smaller $\mathcal{D}_c$, with ±std intervals.}
   \label{fig4}
\end{figure*}

\begin{table*}[t]
\small
\setlength{\tabcolsep}{1.5pt}
\renewcommand{\arraystretch}{0.9}
\centering
\resizebox{1\textwidth}{!}{%
\begin{tabular}{l|cc|cc|cc|cc|cc|cc|cc|c}
\toprule
& \multicolumn{2}{c}{SRS$\rightarrow$H\&E}
& \multicolumn{2}{c}{HO342$\rightarrow$CD3}
& \multicolumn{2}{c}{HO342$\rightarrow$CD8}
& \multicolumn{2}{c}{HE$\rightarrow$ER}
& \multicolumn{2}{c}{HE$\rightarrow$HER2}
& \multicolumn{2}{c}{HE$\rightarrow$PR}
& \multicolumn{2}{c|}{HE$\rightarrow$Ki67} 
& \\
Method
& {\scriptsize Pix2PixHD} & {\scriptsize VSGD+pNCE}
& {\scriptsize Pix2PixHD} & {\scriptsize VSGD+pNCE}
& {\scriptsize Pix2PixHD} & {\scriptsize VSGD+pNCE}
& {\scriptsize CycleGAN} & {\scriptsize CUT}
& {\scriptsize CycleGAN} & {\scriptsize CUT}
& {\scriptsize CycleGAN} & {\scriptsize CUT}
& {\scriptsize CycleGAN} & {\scriptsize CUT} & \textbf{Avg.} \\
\midrule
NHP, $l=1$ 
& 49.3\textsubscript{±5.5} 
& 37.2\textsubscript{±5.7} 
& \color{red}41.0\textsubscript{±3.3} 
& \color{red}46.5\textsubscript{±18.6} 
& \color{red}26.3\textsubscript{±7.9} 
& \color{red}31.0\textsubscript{±14.6} 
& 69.3\textsubscript{±2.9} 
& 62.2\textsubscript{±6.3} 
& 51.0\textsubscript{±5.5} 
& 43.4\textsubscript{±8.9} 
& 68.9\textsubscript{±1.9} 
& 57.8\textsubscript{±4.5} 
& 67.0\textsubscript{±3.6} 
& 47.9\textsubscript{±20.5} & 49.9 \\
NHP, $k=1$ 
& 49.6\textsubscript{±5.4} 
& 37.2\textsubscript{±5.7} 
& 51.1\textsubscript{±4.0} 
& \color{red}47.8\textsubscript{±16.1} 
& \color{red}44.0\textsubscript{±5.1} 
& \color{red}31.6\textsubscript{±14.4} 
& \color{red}66.2\textsubscript{±2.4} 
& \color{red}60.9\textsubscript{±6.6} 
& 51.0\textsubscript{±5.8} 
& 43.2\textsubscript{±8.7} 
& \color{red}67.6\textsubscript{±2.7} 
& 57.1\textsubscript{±5.1} 
& \color{red}65.2\textsubscript{±3.4} 
& 47.0\textsubscript{±20.2} & 51.4 \\
NHP, $\gamma=-1$ 
& \color{red}45.2\textsubscript{±7.9} 
& \color{red}35.8\textsubscript{±7.6} 
& \color{red}40.0\textsubscript{±3.3} 
& 49.4\textsubscript{±16.4} 
& \color{red}25.5\textsubscript{±8.0} 
& 32.6\textsubscript{±16.4} 
& 69.0\textsubscript{±3.1} 
& \color{red}57.5\textsubscript{±5.3} 
& \color{red}49.5\textsubscript{±5.1} 
& 43.3\textsubscript{±8.7} 
& 68.9\textsubscript{±1.9} 
& \color{red}54.8\textsubscript{±5.3} 
& 66.2\textsubscript{±3.3} 
& \color{red}42.2\textsubscript{±19.8} & 48.6 \\
\bottomrule
\end{tabular}}
\caption{Mean (±std.) HRP (\%, $\uparrow$) with NHP hyperparameters fixed at their overall optimum. Declines by +1\% from NHP shown in \color{red}{red}.}
\label{tab2}
\end{table*}

\subsection{Results and Discussion}
\label{5.2}
\subsubsection{Qualitative and Quantitative Comparison} Tab. \ref{tab1} reports the mean HRP using PSNR, SSIM, and LPIPS as $\mathcal{Q}$. NHP outperforms all baselines with HRP significantly above zero (random), suggesting it effectively identifies test images with poor $\mathcal{Q}$ metrics. This performance is consistent across all 14 VS settings, confirming NHP's robustness and ability to generalize to unseen data despite self-tuning. \rev{Note that a perfect (100\%) average HRP is theoretically unattainable here due to ranking disagreements between metrics; e.g., in our work, Kendall's $\tau$ ranged from 0.4--0.5, indicating only moderate rank correlation. Nonetheless, NHP achieves the goal of overall optimal performance.}

Conversely, GAN-based detectors fail, yielding zero or negative HRP. This reinforces concerns reported in previous studies on their sensitivity to unstable training and their OOD-centric detection scope. For instance, ALOCC fails if the discriminator is under- or over-trained \cite{zaheer2020old} and cannot detect realistic hallucinations ($G(\mathbf{s}) \in P_T$) as it only penalizes deviations from the target distribution. While performant in previous studies \cite{ounissi2025scalable, huang2025robust}, their failure here suggests a requirement for careful task-specific GAN tuning beyond our default settings. \rev{DE is more competitive, yielding positive HRP by capturing both aleatoric and epistemic uncertainty via the same $\mathcal{Q}$ metrics used for HRP evaluation. However, it does not surpass NHP. The 4-member setup is unstable and performs poorly, while the 10-member setup improves results but only beats NHP on the SRS$\rightarrow$H\&E task. DE’s limitations may stem from poor individual model calibration \cite{ashukhapitfalls} or an inability to capture bias-driven epistemic uncertainty \cite{lahlou2023deup}, where models ``hallucinate alike'' due to shared inductive biases, resulting in low predictive divergence that escapes detection. Finally, while DE and GAN-based methods incur additional computational overhead from multiple VS training runs or additional model forward passes}, NHP achieves superior performance with only the minimal overhead of a KNN search.

\rev{Visually, Fig. \ref{fig4}-a shows select ROIs from the top and bottom NHP confidence quantiles in the SRS$\rightarrow$H\&E task. High-confidence ROIs exhibit superior fidelity, faithfully reproducing tissue architecture and histological features, such as dense epithelial nuclear clusters and surrounding stromal textures. Conversely, low-confidence ROIs show clear discrepancies: columns 4--5 exhibit significant nuclear loss, while in columns 6--8, regions indicated by arrows are either washed out or replaced by structurally distinct features. These represent a loss of histological context, i.e., hallucinations, which NHP successfully identifies.}

\subsubsection{Justification of NHP's Design} While all NHP variants are performant, our proposed version is most optimal. In Tab. \ref{tab1}, we observe a striking performance drop, sometimes exceeding 20\%, when ablating toward the naive KNN baseline ($\gamma=0$, $q=0$). The linear balance was also effective but slightly underperformed, explaining why it was not chosen. Among distance metrics, KNN yields strongest results, likely due to its non-parametric nature, which better accommodates diverse signal priors and feature distributions in VS datasets. In contrast, other distance metrics impose stronger assumptions--hyperrectangular (OtB), linear (Residual), or Gaussian (GMM)--which may be helpful in specific scenarios but not elsewhere; e.g., Residual performs well on HO342 but fails on SRS.

While this highlights the necessity of key components like FN incorporation, pruning, and KNN use, a remaining question is whether self-tuning is needed, or if a fixed set of hyperparameters, i.e., a universal sweet spot, can perform well across settings. To investigate, we compile the converged hyperparameters across all NHP runs and present their histograms in Fig. \ref{fig4}-b. We observe that the penultimate layer ($l=1$) is generally preferred, though earlier layers ($l \leq 0.25$) sometimes perform better in HO342, while bottleneck layers are rarely optimal. For $q$, aggressive pruning ($q=0.75$) is consistently favored. In contrast, the distributions for $k$ and $\gamma$ are highly dispersed. Overall, NHP does not converge on consistent hyperparameter choices, except for $q$, even within the same VS setting, due to different self-tuning splits. This lack of consistency is further confirmed in Tab. \ref{tab2}, where fixing hyperparameters to their empirically ``best'' values ($l=1$, $k=1$, $\gamma\approx -1$) results in inferior performance. This supports our hypothesis that no single hyperparameter setting is optimal, and grid search is essential for adapting NHP to each task. 

\begin{table*}[t]
\small
\centering
\begin{minipage}[t]{0.48\linewidth}
\centering
\resizebox{\textwidth}{!}{%
\begin{tabular}{l|cc|cc|c}
\toprule
& \multicolumn{2}{c}{HO342$\rightarrow$CD3} & \multicolumn{2}{c|}{HO342$\rightarrow$CD8} & \multicolumn{1}{c}{} \\
Method & {\scriptsize Pix2PixHD} & {\scriptsize VSGD+pNCE} & {\scriptsize Pix2PixHD} & {\scriptsize VSGD+pNCE} & \textbf{Avg.} \\
\midrule
ALOCC & -25.0\textsubscript{±3.7} & 21.8\textsubscript{±22.8} & -28.7\textsubscript{±5.4} & 23.9\textsubscript{±25.4} & -2.0 \\
\rev{DE, 4 mem.} & 23.1\textsubscript{±20.7} & 19.6\textsubscript{±21.9} & 27.1\textsubscript{±23.1} & 13.2\textsubscript{±13.7} & 20.7 \\
\rev{DE, 10 mem.} & \textbf{46.4\textsubscript{±3.9}} & 35.3\textsubscript{±16.1} & 38.4\textsubscript{±7.6} & 11.3\textsubscript{±6.8} & 32.9 \\
NHP, linear & \underline{38.3\textsubscript{±4.7}} & 48.1\textsubscript{±19.3} & \underline{41.6\textsubscript{±8.9}} & \underline{45.9\textsubscript{±22.7}} & \underline{43.5} \\
NHP, GMM & 36.8\textsubscript{±4.1} & \underline{49.1\textsubscript{±21.2}} & 37.0\textsubscript{±7.0} & 44.7\textsubscript{±22.1} & 41.9 \\
NHP & 36.9\textsubscript{±4.0} & \textbf{49.7\textsubscript{±15.6}} & \textbf{42.0\textsubscript{±9.0}} & \textbf{46.0\textsubscript{±22.2}} & \textbf{43.7} \\
\bottomrule
\end{tabular}}
\end{minipage}
\hfill
\begin{minipage}[t]{0.485\linewidth}
\centering
\resizebox{\textwidth}{!}{%
\begin{tabular}{l|cc|cc|c}
\toprule
& \multicolumn{2}{c}{Prov-GigaPath} & \multicolumn{2}{c|}{On-site DINO ViT} & \multicolumn{1}{c}{} \\
Method & {\scriptsize Pix2PixHD} & {\scriptsize VSGD+pNCE} & {\scriptsize Pix2PixHD} & {\scriptsize VSGD+pNCE} & \textbf{Avg.} \\
\midrule
ALOCC & 16.5\textsubscript{±17.8} & 10.4\textsubscript{±15.0} & -26.4\textsubscript{±12.6} & -8.6\textsubscript{±24.1} & -2.0 \\
\rev{DE, 4 mem.} & 29.2\textsubscript{±15.7} & \underline{43.8\textsubscript{±6.2}} & 37.4\textsubscript{±7.8} & 26.7\textsubscript{±8.1} & 34.3 \\
\rev{DE, 10 mem.} & \underline{29.6\textsubscript{±17.3}} & \textbf{46.8\textsubscript{±5.2}} & 37.5\textsubscript{±6.1} & 27.0\textsubscript{±5.3} & \textbf{35.2} \\
NHP, linear & 29.5\textsubscript{±14.9} & 37.1\textsubscript{±7.7} & 29.9\textsubscript{±11.0} & 27.2\textsubscript{±10.1} & 30.9 \\
NHP, GMM & 23.5\textsubscript{±18.2} & 23.6\textsubscript{±9.4} & \underline{37.9\textsubscript{±7.1}} & \textbf{30.6\textsubscript{±13.6}} & 28.9 \\
NHP & \textbf{29.9\textsubscript{±14.7}} & 38.2\textsubscript{±10.1} & \textbf{38.0\textsubscript{±8.8}} & \underline{27.4\textsubscript{±11.6}} & \underline{33.4} \\
\bottomrule
\end{tabular}}
\end{minipage}
\caption{Mean (±std.) HRP (\%, $\uparrow$). \textbf{Left:} HRP using MIR\textsubscript{rel} for HO342 VS tasks. \rev{\textbf{Right:} HRP using structure error of Prov-GigaPath and on-site DINO ViT for the SRS$\rightarrow$H\&E task.} Best in \textbf{bold}, runner-up \underline{underlined}.}
\label{tab3}
\end{table*}

\begin{figure*}[t]
  \centering
  \includegraphics[width=1\linewidth]{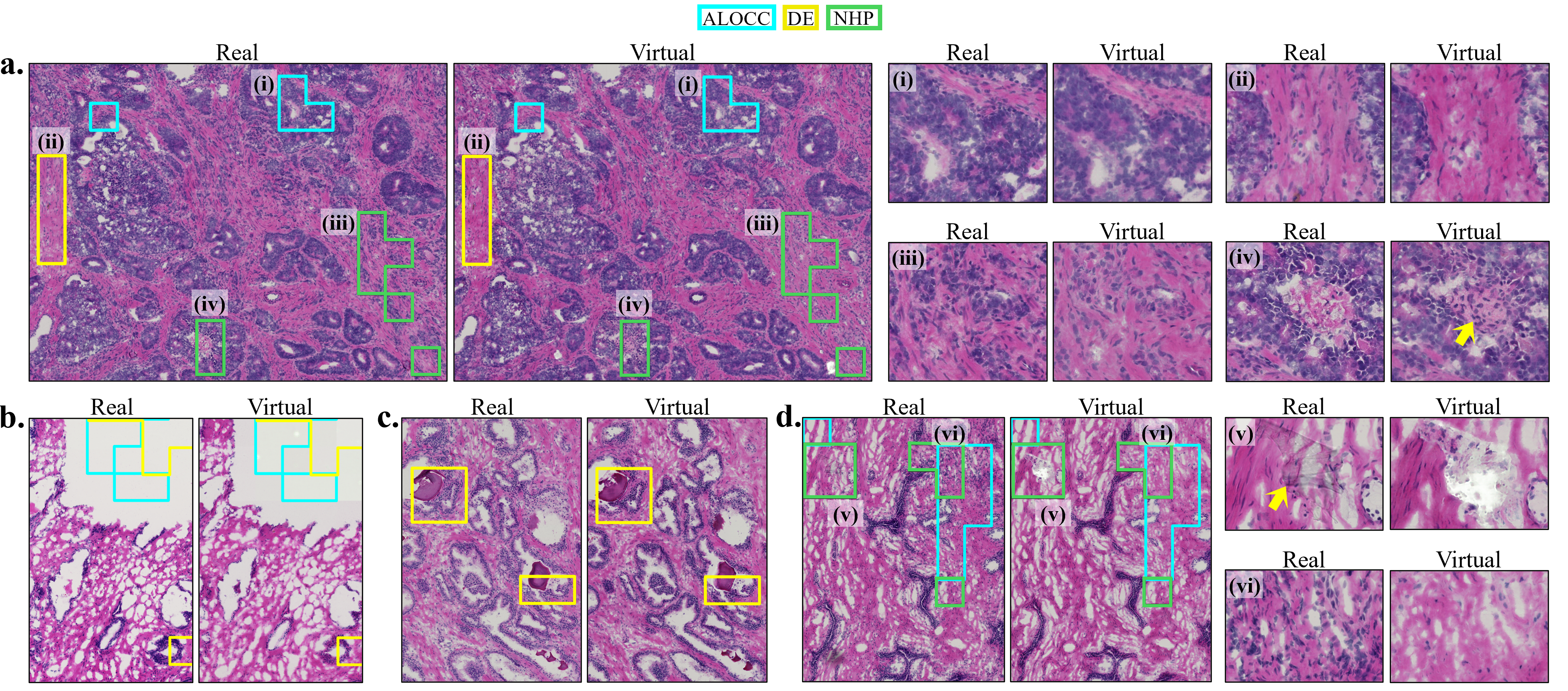}
   \caption{\rev{Selected ROIs from SRS$\rightarrow$H\&E WSIs (\textbf{a}-\textbf{d}) showing ground truth and virtual H\&E images. Top hallucinatory patches (per slide) are marked for NHP and selected methods, with representative zoom-ins (i-vi). Note: since patches are selected per slide, ROIs shown may not contain attended regions for some methods.}}
   \label{fig5}
\end{figure*}

\subsubsection{Sensitivity Analysis} Recall that there is a trade-off between NHP's calibration set size and detector performance. While smaller sets enhance efficiency, they may compromise accuracy, particularly under the aggressive pruning found crucial in previous analyses. While our default sizes (3–7K) are already small, we {test its limits by} evaluating NHP with even fewer samples. In select VS tasks, we downsample $\mathcal{D}_c$ by factors of ${2,4,8,16,32}$ and evaluate HRP in Fig. \ref{fig4}-c, finding most NHPs maintain strong performance down to 100--200 samples. This is notable as the final bank (post validation split and pruning) occasionally contain $N < 50$ samples. SRS was an exception, likely because its smaller training set ($\sim$4K) yields under-regularized latent spaces requiring more samples to mitigate noise. Nevertheless, NHP remains competitive down to 1K samples, demonstrating robustness under tighter resource constraints.

\subsubsection{Extension to Other $\mathcal{Q}$ Metrics} We investigate if NHP remains effective under alternative hallucination definitions ($\mathcal{Q}$) more aligned with clinical utility and pathologist perception:
\begin{itemize}
    \item {Relative masked intensity ratio (MIR\textsubscript{rel})} for HO342: This metric \cite{wolflein2023hoechstgan} assesses how faithfully VS pixel intensities match ground-truth cell segmentation masks.
    \item \rev{{Deep pathology feature error} for SRS$\rightarrow$H\&E: To improve task alignment over LPIPS, which is pre-trained on natural images, we utilize structural error \cite{tumanyan2022splicing} from H\&E-pretrained vision transformers (ViTs) \cite{dosovitskiyimage, amir2021deep}. We evaluate two extractors: Prov-GigaPath \cite{xu2024whole}, a foundation model, and an on-site ViT trained via DINO \cite{caron2021emerging} to minimize site-specific shift.}
\end{itemize}

\color{black}We repeat our experiments by tuning detectors and evaluating HRP using these alternative $\mathcal{Q}$ metrics. Tab. \ref{tab3} shows the results of NHP, baselines, and runner-up methods, where NHP maintains strong, stable performance, generally ranking the first or second. This demonstrates that NHP is $\mathcal{Q}$-agnostic, \rev{which is a significant advantage for real-world applications where users may select hallucination metrics based on clinical context or downstream use. Qualitatively, we compare detected patches in WSIs for the SRS$\rightarrow$H\&E task (Fig. \ref{fig5}). ALOCC and DE occasionally highlight minor pathological deviations, such as nuclear/stromal distributions (a-i, a-ii) or prostatic corpora amylacea (c), and sometimes erroneously flag clinically irrelevant white background (b). Conversely, NHP more consistently pinpoints meaningful hallucinations. For example, (a-iii) and (d-vi) reveal regions with underrepresented nuclei, while (d-v) highlights an OOD artifact in the H\&E ground truth that, despite being diagnostically harmless, causes the VS model to fail catastrophically.}

\begin{figure*}[t]
  \centering
  \includegraphics[width=1\linewidth]{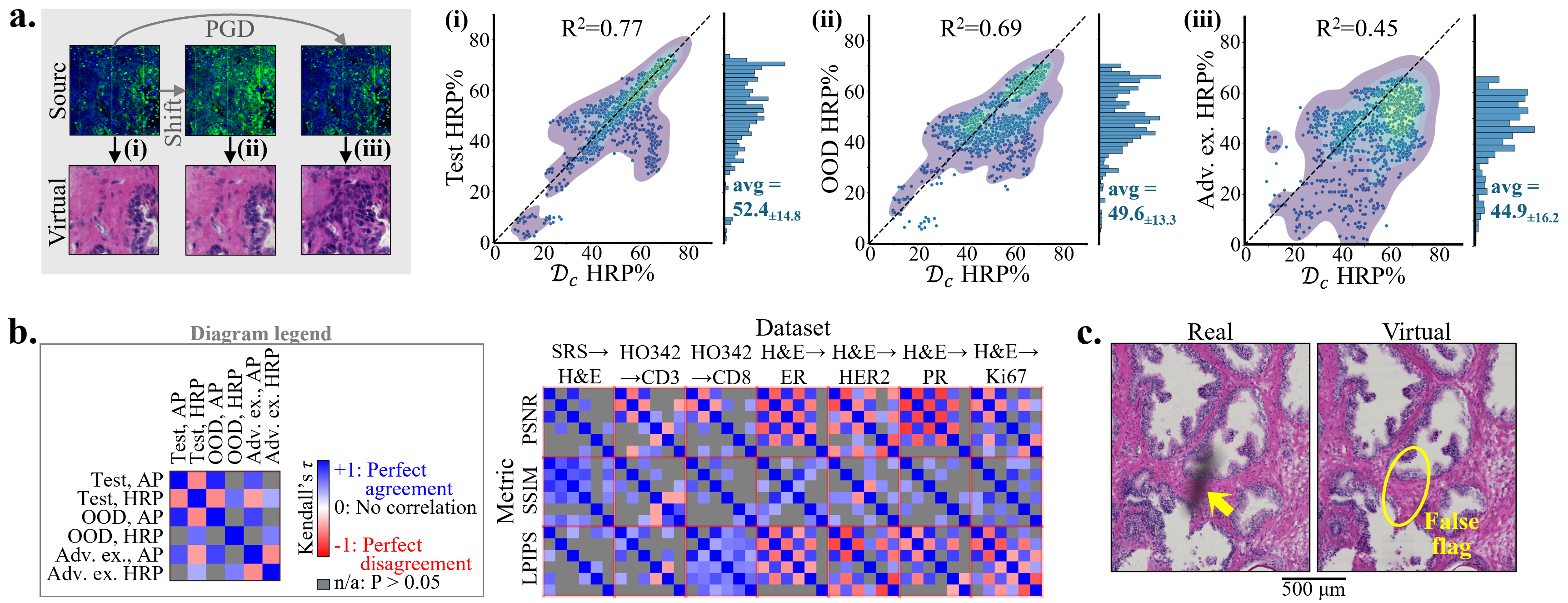}
  \caption{\textbf{a.} Comparison of HRP values on $\mathcal{D}_c$ versus the test, OOD, and adversarial sets. As shown in the left schematic, the VS model is robust to specific shifts but fails under PGD attacks; detection should align with these individual trends. \textbf{b.} Pairwise rank correlations (Kendall's $\tau$) between six safety metrics, computed per $\mathcal{Q}$ across seven VS tasks. \rev{\textbf{c.} Example H\&E target containing a local artifact correctly absent in the virtual image. Although not a hallucination, such regions may be erroneously flagged due to low $\mathcal{Q}$ scores.}}
   \label{fig6}
\end{figure*}

\rev{\subsubsection{Pathologist Validation} While NHP is adaptable to various computational $\mathcal{Q}$ metrics, the ideal benchmark remains expert feedback. However, utilizing pathologist judgment to calibrate NHP is impractical at scale, as it would require experts to manually assign hallucination scores to thousands of individual patches--a process prone to high cost, time constraints, and inter-observer variability. Instead, we evaluate if NHP, when tuned via the previously discussed metrics, aligns with expert judgment. Specifically, we sampled patches from the top and bottom 10\% of NHP scores (based on the on-site DINO ViT structural error) from the SRS$\rightarrow$H\&E task, followed by quality control to ensure patches contained significant glandular and stromal structures. A board-certified pathologist (J.C.) was then asked to blindly evaluate randomized real vs. virtual pairs of high- and low-score patches to determine which appeared more hallucinatory. We assigned a score of +1 for agreement with NHP, -1 for disagreement, and 0 for equivalence. Across 22 trials, NHP achieved a mean score of 0.41, indicating moderate agreement with expert assessment. While a fraction of the disagreement stems from NHP's inability to perfectly mimic $\mathcal{Q}$, another source lies in the limitation of existing computational metrics to fully capture the nuances of pathologist perception. In such cases, the limitation lies in the choice of $\mathcal{Q}$ rather than the NHP formulation itself. Nonetheless, since NHP's success hinges on an effective $\mathcal{Q}$ metric, this highlights a critical need in the broader digital pathology field for image similarity metrics tailored to histopathology.}

\subsubsection{Harder settings}\color{black}The robustness of latent spaces explains NHP's effectiveness on unseen test samples despite self-tuning. Here, we probe the limits of this robustness under more challenging conditions:
\begin{itemize}
    \item {OOD:} Although our test sets are already external, we apply modality-specific corruptions to simulate more severe distribution shifts. For SRS, these include Gaussian noise, contrast jitter, pixel dropout/saturation, band misregistration, and defocus/motion/zoom blurs. For HO342, we apply noise, blurs, contrast shifts, and dropout. For MIST, we introduce JPEG/WebP compression, stain color variation, blurs, and superimposed marker or bubble artifacts \cite{zhang2022benchmarking}.
    \item {Adversarial examples:} We use Projected Gradient Descent (PGD) \cite{madry2018towards}, perturbing each test sample $\mathbf{s}$ into $\mathbf{s}^{\text{adv}}$ to maximize the error between $G(\mathbf{s}^{\text{adv}})$ and either the ground truth $\mathbf{t}$ or the model’s original prediction $G(\mathbf{s})$ (simulating cases where $\mathbf{t}$ is unknown to the attacker). The update rule at step $t$ is:
    \begin{equation}
    \begin{split}
           \textbf{s}^{\text{adv}}_{t+1} = \Pi \Bigl( \textbf{s}^{\text{adv}}_{t} + \alpha \cdot \text{sign}\bigl(\nabla_\textbf{s}(||{G}(\textbf{s}^{\text{adv}}_t) - \xi||_2^2)\bigl)\Bigl),\\s.t. \quad \xi \in \{\textbf{t}, {G}(\textbf{s})\}, \, ||\textbf{s}^{\text{adv}}_{t+1} - \textbf{s}||_{p} \leqslant \epsilon,
    \end{split}
    \label{eq6}
    \end{equation}
    where $\epsilon$ is the perturbation budget, $\alpha$ is the step size, $\nabla_\textbf{s}$ is the gradient w.r.t. $\textbf{s}$, and $\Pi$ projects to the $\ell_p$-ball. We randomly apply either $\ell_2$ or $\ell_\infty$ attack with $\alpha = 0.2$ and $1/255$, respectively, sampling $\epsilon \in \{1/255, 4/255, 8/255\}$ and $\xi \in \{\textbf{t}, {G}(\textbf{s})\}$ per sample. We use 10 PGD steps for HO342 and MIST, and 50 for SRS.
\end{itemize}
These present challenging environments where latent features may be less robust. Note, the detection objective remains unchanged: alignment with $\mathcal{Q}$.

In Fig. \ref{fig6}-a, we compare NHP's HRP on the test, OOD, and adversarial sets against its performance on $\mathcal{D}_c$, the training subset used for tuning. Despite increased difficulty, NHPs generally retain performance, achieving average HRPs of 49.62\% (OOD) and 44.87\% (adversarial). However, the validation gap--the HRP drop relative to $\mathcal{D}_c$--widens under shift: while only 1.42\% on the original test set, it grows to 3.99\% and 8.7\% under corruption and adversarial shifts, respectively. While NHP remains robust in these harder settings, narrowing this gap, perhaps by enriching $\mathcal{D}_c$ with more complex samples, remains a promising future direction.

\subsubsection{Performance vs detectability} \rev{Real-world deployment demands models that are reliable across multiple dimensions \cite{hendrycks2021unsolved}. Specifically, VS models must minimize hallucinations, i.e., exhibit high average performance (AP) defined as $\mathbb{E}_{(\mathbf{s},\mathbf{t})\in\mathcal{D}_\text{test}}[\mathcal{Q}(G(\mathbf{s}),\mathbf{t})]$, while remaining effectively identifiable via monitors (high HRP). Having established NHP as a strong baseline, we investigate the relationship between AP and NHP's HRP: do stronger models with higher AP naturally yield higher HRP? Importantly, these metrics are independent; AP quantifies overall hallucination frequency and severity, whereas HRP measures a monitor's ability to reject hallucination samples, regardless of frequency. For each VS task, we compute six metrics--AP and NHP’s HRP across test, OOD, and adversarial sets--for 20 distinct models (2 backbones $\times$ 10 seeds) and assess the rank correlation for all $\binom{6}{2}$ safety metric pairs (Fig. \ref{fig6}-b). We perform this analysis separately for each $\mathcal{Q}$ metric, as their differing scales preclude averaging operations when computing AP.}

\rev{We observe distinct checkerboard patterns where AP and HRP are often at odds: models with higher AP frequently exhibit lower HRP, and vice versa. This discord sometimes persists even within single metric types, such as HRP\textsubscript{MS-SSIM} for OOD vs. adversarial sets in HO342$\rightarrow$CD3. This suggests that gains in one safety dimension do not guarantee, and may even undermine, performance in another, echoing ML literature where models with higher accuracy often exhibit poorer precision-recall, calibration, or adversarial robustness \cite{tsipras2019robustness, chun2020empirical}. Specifically, we speculate this negative AP vs. HRP correlation may be linked to ``feature collapse'' \cite{van2021feature, mukhoti2023deep}. In GANs, anomalous latent features in the extreme tails often trigger artifact synthesis \cite{brock2018large, song2023feature}; stronger VS models may mitigate these by producing more tightly bounded latent spaces. However, this collapsed density may inadvertently degrade separability between hallucination and non-hallucination features for NHP. While this indicates a limitation of NHP and latent-space approaches, other GAN-based and DE approaches did not yield superior HRP. Thus, an all-around reliable VS model remains elusive, revealing a critical gap in the literature. While recent research focuses on improving test-set AP \cite{liu2021unpaired, zeng2022semi, lin2022unpaired, boyd2022region, li2023adaptive, ma2023efficient, li2024virtual, chen2024pathological, wang2024mix, wei2024derestainer}, there is a vital need to integrate hallucination detection into standard VS benchmarks, as optimizations for AP may inadvertently compromise detectability.}

\subsection{Limitations \& Future Work}
\label{sec:5.3}
Finally, we outline some limitations of this work, inspiring future directions:

\rev{\textit{Algorithmic refinements:} The current NHP represents a baseline formulation, intended as a starting point. Its performance could likely be improved with better distance functions, such as graph-based methods or manifold learning for high-dimensional latent spaces. The hyperparameter sweep space also merits refinement. For instance, we currently sweep over layers using sequential index, assuming it spans diverse representations, but this is not guaranteed. A more principled approach could involve computing inter-layer similarity and selecting a maximally diverse subset of layers.}

\rev{\textit{Target artifact bias:} An unstated assumption in this work is that the target image reflects pristine ground truth. However, real-world histology datasets contain artifacts, which becomes problematic when they appear in the target image but not the source (Fig. \ref{fig5}-c), biasing $\mathcal{Q}$ scores downward. If such pairs enter the safe set, NHP may mistakenly treat them as detection targets despite good VS quality. This highlights the importance of quality control protocols or artifact-robust $\mathcal{Q}$ metrics.}

\textit{Finer source disentanglement:} While we currently utilize a single score to represent all hallucination sources (Fig. \ref{fig2}-a), unmixing them could support targeted clinical interventions to minimize specific uncertainties. For example, hallucinations from epistemic factors demand active learning\rev{--i.e., incorporating the tissue into the training set after staining to improve VS model performance}. In contrast, aleatoric sources are irreducible and indicate issues within the source modality itself, such as low content. Certain cases, like microscopy artifacts, may even be resolved by re-scanning the source image. Furthermore, subtyping hallucinations by threat level (e.g., minor vs. catastrophic) could better assist pathologists in triaging diagnostic risks.

\textit{Finer spatial attribution:} We adopt patch-level confidence, which aligns with WSI-scale analysis and the interpretability resolution of common frameworks like MIL \cite{ilse2018attention, kapse2024si}. However, finer resolution may be preferred for VS with downstream ``needle-in-a-haystack'' applications, such as detecting micro-metastases, \rev{isolated tumor cells, or mitoses}.

\rev{\textit{Large-scale validation:} Due to the lack of public large-scale VS datasets, validation is limited to relatively small, single-site cohorts. While results show promise, further studies are needed. For instance, scaling behaviors seen with small calibration set sizes may not hold for larger, more heterogeneous, and long-tailed datasets, demanding larger calibration sets. Moreover, clinical deployment may involve running multiple VS models in parallel for multi-stain tasks, imposing stricter memory and runtime constraints per model and further increasing the challenge. In such scenarios, approximate KNN techniques like Faiss's IVF and IVFPQ may be necessary.}

\rev{\textit{Unified mitigation and detection:} This work focuses on post-hoc hallucination detection, while a more mainstream, orthogonal strategy targets mitigation during training. Both are essential for trustworthy VS, yet they do not naturally align, as we showed. A promising direction is to develop unified VS frameworks that reduce hallucinations while also make them easier to detect. Similar goals are gaining traction in the broader ML community \cite{bai2023feed}.}

\section{Conclusions}
In this work, we studied hallucination detection for VS, contributing in three key dimensions. First, we formally established the problem, providing necessary background and clarification. \rev{Specifically, we highlighted the diverse sources of hallucinations across the VS train-deploy pipeline and argued that detection strategies should align with this complexity, rather than rely on proxy tasks like OOD or outlier detection.} Second, we introduced a baseline method, NHP, with rigorous validation. \rev{We demonstrated its simplicity (as an extension of KNN), versatility (agnostic to dataset, I2IT model, or hallucination metric $\mathcal{Q}$), robustness (effective under harder settings), and scalability (low computational overhead).} Third, we uncovered new insights into VS robustness, \rev{notably that models with fewer hallucinations do not necessarily exhibit better detection.}

\rev{Broadly, our study is motivated by the growing interest in VS across biomedical research and clinical landscapes. While recent advances are bringing us closer to hyper-realistic VS outputs, no AI model is immune to failure. Given the high-stakes nature of digital pathology, where hallucinations may carry adverse clinical consequences, there is a pressing need to study hallucination detection as an essential next step toward trustworthy VS deployment. In this regard, we hope our work serves as a primer for VS researchers and practitioners.}

\bibliographystyle{ieeetr}
\bibliography{references}
\end{document}